# Extension of the composite CBS-QB3 method to singlet diradical calculations

Baptiste Sirjean [a], René Fournet [a], Pierre-Alexandre Glaude [a], Manuel F. Ruiz-Lopez [b*]

[a] *Département de Chimie Physique des Réactions, UMR 7630 CNRS-INPL, Nancy-Université 1, rue Grandville, BP 20451, 54001 Nancy Cedex, France*
[b] *Equipe de Chimie et Biochimie Théoriques, UMR CNRS-UHP No. 7565, Nancy-Université, Boulevard des Aiguillettes, BP 239, 54506 Vandoeuvre-lès-Nancy, France*

[*] Corresponding author. Fax: +33 3 83 68 43 71
*E-mail address:* Manuel.Ruiz@cbt.uhp-nancy.fr

**Abstract**

The composite CBS-QB3 method is widely used to obtain accurate energies of molecules and radicals although its use in the case of singlet diradicals gives rise to some difficulties. The problem is related to the parameterized correction this method introduces to account for spin-contamination. We report a new term specifically designed to describe singlet diradicals separated by at least one $CH_2$ unit. As a test case, we have computed the formation enthalpy of a series of diradicals that includes hydrocarbons as well as systems involving heteroatoms (nitrogen, oxygen). The resulting CBS-QB3 energies are very close to experiment.

## 1. Introduction

The accurate computation of thermochemical quantities requires the use of elaborated quantum mechanical methods accounting for both static and dynamic electron correlation effects. Unfortunately, such methods (MRCI for instance) are very costly and are therefore confined to small system investigations. Composite methods have been developed with the aim of reaching chemical accuracy in larger systems. One of such methods is the CBS-QB3 approach that belongs to the family of the Complete Basis Set (CBS) methods of Petersson and co-workers [1]. It shows mean absolute deviation of only 1.1 kcal/mol on the G2/97 set test [2] and not surprisingly, its popularity is rapidly increasing.

In CBS-QB3 calculations, the following steps are involved. Optimization and frequency calculations are performed at the B3LYP/CBSB7 level [3-4]. Afterwards, single point calculations are performed at CCSD(T)/6-31+G(d') and MP4SDQ/CBSB4 levels. The total energy is extrapolated to the infinite-basis-set limit using pair natural-orbital energies at the MP2/CBSB3 level and an additive correction to the CCSD(T) level. In addition, CBS-QB3 includes a correction for spin contamination in open-shell species, namely

$$\Delta E_{spin} = -0.00954 \; (<S^2> - <S^2_{th}>) \quad (1)$$

where $<S^2>$ denotes the actual eigenvalue of the $S^2$ operator at the MP2/CBSB3 level of calculation and $<S^2_{th}>$ the corresponding theoretical value (e.g. 0 for a singlet, 2 for a triplet, etc). The constant $-0.00954$ in equation (1) was optimised using eight reference values: the C-H dissociation energies of HCN, $C_2H_4$ and $C_2H_2$, the ionisation potentials of CS and CO, and the electronic affinities of CN, NO and PO [1].

CBS-QB3 is based on the use of one determinantal wavefunctions and therefore the description at this level of open-shell singlet diradicals (OSD) may be questioned. As far as geometry optimization is concerned, however, this appears to be a minor problem. Indeed, recent papers have shown that for OSD with separated radical centers, geometries obtained at the unrestricted-DFT (UDFT) approach compare well with those obtained at more refined levels [5-7]. The UDFT method allows spin symmetry to be broken, i. e. α and β orbitals are permitted to have different spatial functions. Spin-contamination, on the contrary, is a much more delicate problem. One-determinantal wave functions for OSD usually lead to $<S^2>$ values close to 1 owing to an almost equivalent mixture of singlet and triplet states. Such a strong spin contamination is not correctly handled by equation (1) in the CBS-QB3 method and a systematic error of about 6 kcal/mol is introduced by this equation. In a recent paper [8], a modification of the standard CBSQB3 method was introduced to overcome the problem of misleading empirical correction for severely spin-contaminated species. The authors have proposed a spin-restricted open-shell complete basis set model chemistry (ROCBS-QB3) to eliminate spin contamination for any open-shell species. The ROCBS-QB3 model shows good reliability for the set of species studied but suffer from the own problems of spin-restricted procedures and unfortunately cannot treat widely separated diradical centers.

Such diradicals are involved in many important gas-phase and heterogeneous industrial processes such as combustion, partial oxidation, cracking, pyrolysis or photochemical reactions. Direct experimental studies of diradicals have been made by Zewail and coworkers using femtosecond laser techniques, that allow establishing the nature of these species [9]. It appears therefore necessary to investigate how the CBS-QB3 method could be modified in order to overcome the difficulty mentioned above.

In this paper, we propose a term to account for spin contamination in singlet diradical CBS-QB3 calculations. The present approach focuses on diradicals separated by at least one $CH_2$ unit since in the opposite case, the CBS-QB3 methodology is not expected to work. The correction term contains a single parameter that has been optimized so that CBS-QB3 reproduces the singlet-triplet gap obtained in CASSCF computations for twelve hydrocarbons diradicals. As a test case, we have computed the formation enthalpy of 22 diradicals and we compare the results with experimental data.

## 2. Calculation details

Calculations were performed with Gaussian 03 Rev. B.05 [10] for all diradicals. Geometry optimizations and frequencies calculations were performed at the B3LYP/CBSB7 level for both singlet and triplet states. For OSD computations, the keyword GUESS=MIX was used in order to break the α/β symmetry. Note that the standard route for CBS-QB3 calculations has to be modified in order to be able to use the option GUESS=MIX. Energies were obtained using the CBS-QB3 procedure, as explained below. Spin densities were systematically inspected in order to verify that we obtained the desired diradicals. Reference singlet-triplet gaps, including geometry optimisation, were calculated at the CASSCF(4,4)/6-311G(d,p) level. In the latter calculations, the eigenvectors of the CI matrix were inspected in order to check the correctness of singlet and triplet state configurations.

## 3. Parameterization procedure

In OSD calculations using one-determinantal wave function, singlet-triplet mixing is very large and quite often one assumes the sum method to derive the energy of the pure singlet state provided one knows that of the triplet state [11-12]:

$$E_{OSD} = xE_S + (1-x)E_T \quad (2)$$

Here, $E_{OSD}$ is the energy for the one-determinant OSD calculation, $E_T$ and $E_S$ are the energies of the triplet and singlet states respectively and $x$ is determined from the equation:

$$\langle \hat{S}^2 \rangle_{OSD} = x\langle \hat{S}^2 \rangle_S + (1-x)\langle \hat{S}^2 \rangle_T \quad (3)$$

In many instances, and in particular for hydrocarbon diradicals, $\langle \hat{S}^2 \rangle_{OSD}$ is close to 1 and $\langle \hat{S}^2 \rangle_T$ is close to 2, so that $x$ is close to 0.5. This leads to the so-called 50:50 approximation that we assume in our work:

$$E_S = 2E_{OSD} - E_T \quad (4)$$

Let us now consider a CBS-QB3 calculation for an OSD. The standard energy $E_{OSD}^{CBSQB3}$ contains the correction defined by equation (1). In our scheme, this correction is removed and replaced by a term that is assumed to correct for the singlet-triplet 50:50 energy. Thus, the new $E_{OSD}^{CBSQB3}$ energy is given by:

$$E_{OSD}^{CBSQB3} = E_{OSD,uncorr}^{CBSQB3} + Y(\langle \hat{S}^2 \rangle_{OSD} - 1) \quad (5)$$

where $E_{OSD,uncorr}^{CBSQB3}$ is the CBS-QB3 energy without the correction given by equation (1) and $Y$ is a parameter to be determined. The pure singlet energy can then be estimated from equation (4):

$$E_S^{CBSQB3} = 2E_{OSD}^{CBSQB3} - E_T^{CBSQB3} \quad (6)$$

and the singlet-triplet energy gap can be computed from:

$$\Delta E_{S-T}^{CBSQB3} = E_{S}^{CBSQB3} - E_{T}^{CBSQB3} \qquad (7)$$

or using equation (6):

$$\Delta E_{S-T}^{CBSQB3} = 2(E_{OSD}^{CBSQB3} - E_{T}^{CBSQB3}) \qquad (8)$$

Equations (5) and (8) provides a method to obtain $Y$ by requesting $\Delta E_{S-T}^{CBSQB3}$ to be as close as possible to the singlet-triplet energy gap calculated at the CASSCF level $\Delta E_{S-T}^{CASSCF}$ for some reference compounds. In other words, one has to look for an $Y$ value minimising the following quantity:

$$\Delta E_{S-T}^{CASSCF} - 2(E_{OSD}^{CBSQB3} - E_{T}^{CBSQB3}) \qquad (9)$$

The reference OSDs studied in this paper are summarized in Figure 1. The optimal value of the empirical coefficient in equation (5) has been found to be $Y = -0.031$.

Table 1 and Figure 2 compare the single-triplet energy gap obtained with each one of the following approximations:

(a) The sum method is not applied to obtain singlet OSD energies and the standard CBS-QB3 correction is employed to correct for spin-contamination (both in singlet and triplet states).

(b) Singlet energies are obtained by the sum method (equation 6) and using $Y = 0$ in equation 5, triplet energies are obtained with the standard CBS-QB3 method

(c) Same than (b) but using the optimized value $Y = -0.031$

Table 1 also includes the $\langle \hat{S}^2 \rangle_{OSD}$ values obtained in the CBS-QB3 calculation, which are very close to 1 for all diradicals.

As expected, singlet-triplet gaps obtained with approximation (a) are systematically overestimated with respect to CASSCF energy gaps due to an excessive spin-correction contribution (singlets are too stable). Approximation (b), which has been used by some authors [13-14] in the case of strongly spin-contaminated doublet, improves the absolute value of the energy gap but tends to invert the relative order of stability between singlets and triplets (triplets too stable). Use of the empirical parameter proposed here (approximation (c)) substantially improves the calculations: in all cases, the predicted energy gap is closer to the CASSCF value. It should be noted that for systems exhibiting a slight singlet-triplet gap, the magnitude of the spin-contamination correction for the singlet can be larger than the corresponding correction coming from the sum method. In those cases, $E_{OSD}^{CBSQB3}$ in equation 5 would represent a good approximation for the singlet state energy.

It can be also remarked that the largest error decreases with k, the number of $CH_2$ groups separating the radical centers (but note that the training set includes only one radical with k=4 and one with k=5). These errors amounts roughly 1.1, 0.9, 0.6 and 0.6 kcal/mol for n=2,3,4 and 5 respectively.

## 4. Test case: formation enthalpy of diradicals

We now test the accuracy of the parameter proposed above by comparing CBS-QB3 formation energy computations with experimental estimations. A usual method to estimate the formation enthalpy of a diradical makes use of measured bond dissociation energies (BDE). For $C_nH_{2n+2}$ diradicals in Figure 1, one considers the process:

$$C_nH_{2n+2} \rightarrow \bullet C_nH_{2n}\bullet + 2\ H\bullet \quad (10)$$

where $C_nH_{2n+2}$ stands for a linear or branched alkane and $\bullet C_nH_{2n}\bullet$ for the corresponding diradical. The reaction enthalpy is twice a C-H BDE so that the enthalpy of formation of $\bullet C_nH_{2n}\bullet$ can be calculated, at 298K, from:

$$\Delta H_{f(C_nH_{2n})} = 2BDE + \Delta H_{f(C_nH_{2n+2})} - 2\Delta H_{f(H)} \quad (11)$$

Equivalent expressions can be derived for other diradicals, like those in **Figure 3**, that have been chosen for complementary tests. The BDE method allows a good estimation of the formation enthalpy of diradicals provided no interaction exists between the two radical centers. The use of this method to estimate the enthalpy of formation of diradical species has allowed us to correctly model experimental results in the thermal decomposition of polycyclane [15]. The BDE for primary, secondary or tertiary C–H bonds in linear or branched alkanes are taken from Luo [16]. Formation enthalpies for alkanes and H• are taken from references [17-20].

Theoretical computation of formation enthalpies usually employs isodesmic reactions [21]. When the conservation of the total number and types of bonds applies, there is a large cancellation of errors and very accurate results can be obtained. Several isodesmic reactions have been considered for the calculation of $\Delta H_f°$ in order to get an average value. $\Delta H_f°$ has been obtained for the prototype reactions:

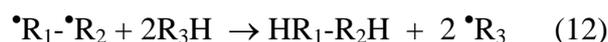

$$\bullet R_1\text{-}\bullet R_2 + 2R_3H \rightarrow HR_1\text{-}R_2H + 2\,\bullet R_3 \quad (12)$$

where $\bullet R_1\text{-}\bullet R_2$ represents a biradical and $\bullet R_3 = \bullet H$, $\bullet C_2H_5$ or n-$\bullet C_3H_7$.

In Table 2, we compare CBS-QB3 calculations and experimental results using the BDE method for the formation enthalpies. The Table includes results for the $C_nH_{2n}$ diradicals in the training set (Figure 1) as well as results for diradicals not in the training set (Figure 3). The latter involve hydrocarbons and heteroatomic systems. An excellent agreement is obtained in all cases. Mean absolute deviation (MAD) for calculated *vs* experimental formation enthalpies amounts only 1 kcal/mol (for diradicals in the training set) or 1.5 kcal/mol (for other diradicals). Note that results for heteroatomic diradicals are very satisfying and that results for diradicals separated by a single $CH_2$ group (17, 20) do not present larger errors than the others. The diradical $\cdot CH_2$-$CH_2$-$CH_2\cdot$ could not be computed since it was found to be unstable (it leads to cyclopropane). Singlet diradicals on the same atom (for instance $CH_2$) or in linked atoms (for instance $\cdot CC\cdot$ EXEMPLE OK ?) were not tested. The present method is not expected to be accurate in those cases because of the strong unpaired electron interactions. On the contrary, results for diradicals with larger separations than those considered here are expected to be at least as good as those reported in Table 2 since the singlet-triplet gap (and therefore errors associated with the 50:50 approximation) should decrease. Finally, it is worth mentioning that extension of the CBS-QB3 method described here has allowed us to compute kinetics parameters for the ring opening of cycloalkanes [22] that are in very good agreement with experimental data.

## 5. Conclusion

In this work, a term to correct the total energy of a singlet diradicals in CBS-QB3 calculations due to spin-contamination has been reported. The approach concerns diradicals separated by at least one $CH_2$ unit. The correction $\Delta E_{spin} = -0.031\ (<S^2> - 1)$ is designed for diradicals described within the broken symmetry approach. The correction term has been established for hydrocarbon diradicals but further tests have allowed us to verify its validity for diradicals involving heteroatoms. Our modified CBS-QB3 method allows us to reproduce the singlet-triplet energy gap calculated with a multi-reference method such as CASSCF as well as the formation enthalpy of diradicals. The modification considerably expands the range of applicability of the CBS-QB3 method, in particular to species involved in important chemical reactions occurring in the atmospheric or in combustion processes.

**Acknowledgements**

The authors thank Christopher Cramer and Josep Anglada for useful comments. The Centre Informatique National de l'Enseignement Supérieur (CINES) is gratefully acknowledged for allocation of computational resources.

**Table 1.** Calculated singlet-triplet energy gaps in kcal/mol at the CASSCF(4,4)/6-311G(d,p) and CBS-QB3 levels for $C_nH_{2n}$ diradicals in Figure 1. CBS-QB3 calculations assume different approximations for the singlet state, as defined in the text. Values of $\langle \hat{S}^2 \rangle_{OSD}$ correspond to the CBS-QB3 calculation. k is the number of $CH_2$ groups separating the radical centers.

| Diradical | n | k | $\Delta E_{S-T}^{CASSCF}$ | $\Delta E_{S-T}^{CBSQB3}$ | | | CBS-QB3 $\langle \hat{S}^2 \rangle_{OSD}$ |
|---|---|---|---|---|---|---|---|
| | | | | (a) Standard | (b) Y=0 | (c) Y=-0.031 | |
| 1 | 5 | 3 | -1.26 | -5.87 | 0.48 | -0.33 | 1.0210 |
| 2 | 5 | 2 | -0.48 | -5.67 | 0.92 | -0.03 | 1.0244 |
| 3 | 6 | 3 | -0.61 | -5.89 | 0.45 | -0.38 | 1.0221 |
| 4 | 6 | 2 | -0.35 | -5.35 | 1.59 | 0.56 | 1.0264 |
| 5 | 6 | 2 | 0.00 | -5.13 | 2.03 | 0.98 | 1.0269 |
| 6 | 6 | 3 | -0.16 | -5.83 | 0.58 | -0.30 | 1.0226 |
| 7 | 5 | 2 | -1.18 | -6.54 | 0.38 | -0.11 | 1.0125 |
| 8 | 6 | 2 | -2.42 | -7.25 | -2.43 | -2.71 | 1.0071 |
| 9 | 4 | 2 | -0.96 | -5.82 | 0.51 | -0.08 | 1.0152 |
| 10 | 6 | 4 | -0.19 | -5.45 | 1.37 | 0.37 | 1.0258 |
| 11 | 7 | 5 | 0.39 | -5.13 | 2.02 | 0.99 | 1.0266 |
| 12 | 6 | 3 | -0.47 | -5.60 | 1.06 | 0.12 | 1.0242 |

**Table 2**. Comparison between the experimental formation enthalpy (obtained by the BDE method) and theoretical calculations at CBS-QB3 level (obtained from isodesmic reactions) for diradicals in Figure 1 (training set) and Figure 3 (complementary test set). Diradicals 17-22 involve heteroatoms. Values in kcal.mol$^{-1}$, at 298 K.

| Diradicals (training set) | Formation enthalpy | | Diradicals (complementary test set) | Formation enthalpy | |
|---|---|---|---|---|---|
| | exp | this work | | exp | this work |
| 1 | 62.6 | 62.3 | 13 | 104.9 | 103.8 |
| 2 | 60.8 | 59.5 | 14 | 55.2 | 56.7 |
| 3 | 55.9 | 56.1 | 15 | 49.0 | 49.4 |
| 4 | 54.3 | 53.2 | 16 | 71.4 | 69.7 |
| 5 | 54.9 | 53.8 | 17 | 86.9 | 87.1 |
| 6 | 56.6 | 55.4 | 18 | 82.5 | 81.0 |
| 7 | 60.7 | 60.5 | 19 | 75.6 | 76.1 |
| 8 | 55.1 | 52.6 | 20 | 45.5 | 46.4 |
| 9 | 67.6 | 67.8 | 21 | 36.0 | 36.5 |
| 10 | 57.6 | 58.3 | 22 | -0.3 | 1.7 |
| 11 | 52.7 | 53.7 | | | |
| 12 | 54.7 | 54.8 | | | |
| *AVE error* | | *-0.5* | | | *0.2* |
| *MAD error* | | *1.0* | | | *1.5* |
| *RMS error* | | *1.0* | | | *1.2* |

**Figure Captions**

**Figure 1.** Reference diradicals considered in our calculations

**Figure 2.** Comparison of singlet-triplet energy gaps obtained at the CBS-QB3 level assuming different approximations for the singlet state, as defined in the text.

**Figure 3**. Diradicals chosen for test calculations in Table 3.